\documentclass[journal]{IEEEtran}

\ifCLASSINFOpdf
\usepackage{graphicx}   
\usepackage{float} 
\usepackage[caption=false]{subfig} 
\usepackage{float} 

\else
\fi
\usepackage{amsmath} 
\usepackage{amssymb} 
\usepackage{tikz} 
\usepackage{tikz} 
\usepackage{hyperref} 
\usepackage{tikz} 
\usetikzlibrary{positioning} 
\usepackage{float}
\begin{document}
%
\title{Seismic Velocity Inversion from Multi-Source Shot Gathers Using Deep Segmentation Networks: Benchmarking U-Net Variants and SeismoLabV3+}
%
%
%
\author{Mahedi Hasan,~\IEEEmembership{Member, IEEE}, Associate Member (AACS), Australian Computer Society}

%

%
%

\markboth{Journal of \LaTeX\ Class Files,~Vol.~13, No.~9, September~2014}%
{Shell \MakeLowercase{\textit{et al.}}: Bare Demo of IEEEtran.cls for Journals}
%



\maketitle

\begin{abstract}
Seismic velocity inversion is a key task in geophysical exploration, enabling the reconstruction of subsurface structures from seismic wave data. It is critical for high-resolution seismic imaging and interpretation. Traditional physics-driven methods, such as Full Waveform Inversion (FWI), are computationally demanding, sensitive to initialisation, and limited by the bandwidth of seismic data. Recent advances in deep learning have led to data-driven approaches that treat velocity inversion as a dense prediction task. This research benchmarks three advanced encoder–decoder architectures—U-Net, U-Net++, and DeepLabV3+—together with SeismoLabV3+, an optimised variant of DeepLabV3+ with a ResNeXt50\_32x4d backbone and task-specific modifications—for seismic velocity inversion using the ThinkOnward 2025 Speed \& Structure dataset, which consists of five-channel seismic shot gathers paired with high-resolution velocity maps. Experimental results show that SeismoLabV3+ achieves the best performance, with MAPE values of 0.03025 on the internal validation split and 0.031246 on the hidden test set as scored via the official ThinkOnward leaderboard. These findings demonstrate the suitability of deep segmentation networks for seismic velocity inversion and underscore the value of tailored architectural refinements in advancing geophysical AI models.

\end{abstract}

\begin{IEEEkeywords}
 Seismic velocity inversion, Deep learning, U-Net, U-Net++, DeepLabV3+, Encoder–decoder networks, Regression-based inversion, SeismoLabV3+, Subsurface imaging
\end{IEEEkeywords}

%
\IEEEpeerreviewmaketitle

\section{Introduction}
\IEEEPARstart{S}{eismic} velocity is a fundamental parameter in seismic
exploration and plays a critical role in subsurface characterization. Accurate velocity models are essential prerequisites
for advanced imaging techniques, such as reverse-time migration (RTM) and other high-resolution seismic processing methods\cite {yang2019deep}. It is not only used for the search for economically valuable resources such as oil, gas and minerals, but also plays a vital role in engineering, archaeological, and geoscientific studies\cite{sciencedirect_seismics}. 

Seismic Conventional physics-driven methods include ray travel
time tomography \cite{hole1992nonlinear}, wave equation tomography\cite{woodward1992wave}, migration velocity analysis \cite{liu1995migration}, \cite{liu2008migration}, and full waveform inversion (FWI) \cite{pratt1998gauss} estimate subsurface velocities by solving forward and adjoint wave equations simulating seismic wave propagation.
These methods rely on iterative optimization to reduce the mismatch between observed and synthetic data. They have several limitations, such as they are computationally intensive and time-consuming, and susceptible to the choice of initial velocity model, which often gets trapped in local minima, and require regularization techniques to address the ill-posed nature of the inversion problem. \cite{wei2021seismic}. 

In contrast, data-driven deep learning (DL) methods aim to learn a direct mapping from seismic shot gatherings to subsurface velocity models without relying on explicit physical modeling. As a powerful non-linear function approximator, DL has been increasingly used to estimate velocity fields efficiently and accurately. Recent advances have led to the emergence of deep learning–based inversion techniques, particularly those that utilise convolutional neural networks (CNNs) and transformer-based architectures, which have gained significant interest in seismic exploration research. These approaches focus on training models to infer velocity structures directly from raw seismograms, offering a promising alternative to traditional physics-driven methods. \cite{wang2023seismic}. Even with these advances, comparative evaluations are still fragmented. Most research focuses on a single architecture within task-specific contexts, which hinders the ability to evaluate performance comparatively under uniform conditions. U-Net, initially proposed for biomedical image segmentation with limited annotated data\cite{UNet}, and DeepLab\cite{chen2017deeplab}, developed for semantic segmentation in natural images, are both highly adaptable. This research work demonstrates their versatility by adapting both architectures, along with their advanced variants, for the seismic velocity inversion task.

The main contributions of this work are summarized as follows:

\begin{enumerate}
    \item \textbf This research benchmarks U-Net, U-Net++, and SeismoLabV3+ (an optimized DeepLabV3+) for seismic velocity inversion, ensuring fairness by applying identical preprocessing steps, hyperparameter tuning protocols, training conditions, and evaluation metrics. 

    \item \textbf This research introduces SeismoLabV3+, a task-adapted variant of DeepLabV3+ modified for regression with a resnext50\_32x4d backbone, five-channel seismic inputs, and task-specific optimizations, which achieves superior performance compared to U-Net and U-Net++ 
\end{enumerate}

These findings highlight the suitability and potential of deep segmentation networks for seismic velocity inversion, demonstrating their ability to capture multiscale geological structures and sharp velocity boundaries, thereby bridging the gap between traditional physics-driven methods and efficient, data-driven approaches.

\section{Problem Statement}
\begin{figure}[H]
\centering
\includegraphics[width=0.48\textwidth]{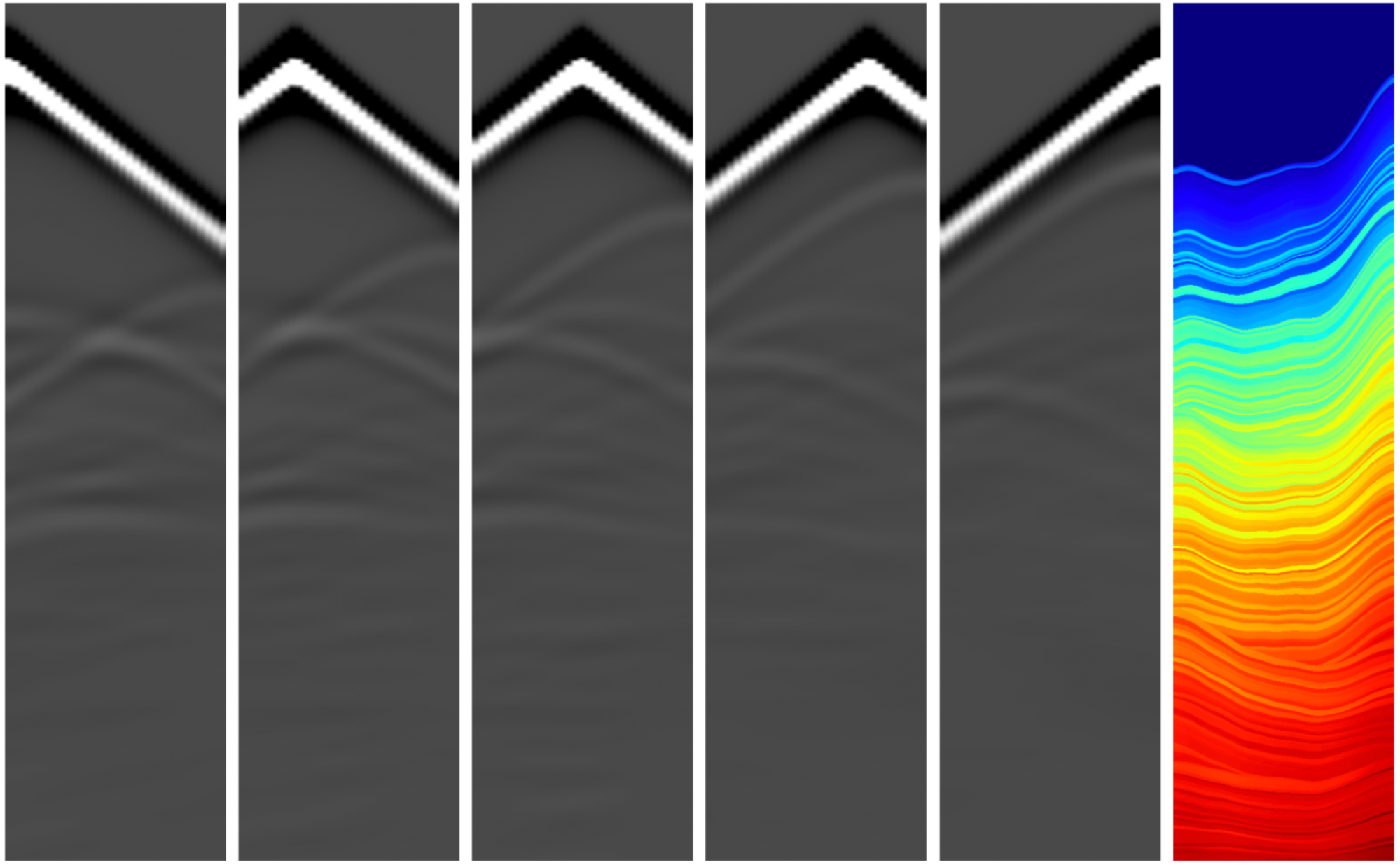}
\caption{Example from the Speed and Structure Challenge: multiple seismic shot gathers (left) with their corresponding ground-truth velocity model (right)\cite{thinkonward2025}.}
\label{fig:speed_structure_overview}
\end{figure}

Seismic velocity inversion refers to the process of reconstructing a velocity model of subsurface structures from seismic shot records, which capture the echoes of sound waves traveling through different layers of the Earth ~\ref{fig:speed_structure_overview}. This velocity information is crucial for representing geological information, identifying natural resources, and directing drilling operations. 

Traditional physics-driven approaches, notably Full Waveform Inversion (FWI), have long been considered the benchmark for seismic velocity inversion because of their ability to generate high-resolution subsurface models. These methods suffer from several inherent limitations, including susceptibility to cycle-skipping, where small misalignments between observed and simulated waveforms can lead to convergence on incorrect models. Additionally, large-scale inversions are computationally intensive, often requiring substantial processing time and necessitating high-performance computing facilities. Their effectiveness depends heavily on the accuracy of the initial velocity model supplied. The band-limited nature of seismic acquisition constrains the recoverable resolution, further compounding the ill-posedness of the inversion problem. \cite{pratt1998gauss, virieux2009, brossier2009seismic, thinkonward2025}. To address these challenges, recent research has shifted towards data-driven approaches that cast seismic inversion as a supervised pixel-wise regression problem, allowing neural networks to directly learn the mapping from seismic shot gatherings to subsurface velocity models. \cite{yang2019deep, wu2019inversionnet}.
Instead of iteratively solving wave equations, deep neural networks are trained to map seismic shot gatherings to subsurface velocity models directly. Formally, this mapping can be mathematically described as:

\begin{align}
    \text{Input:} \quad & \mathbf{X} \in \mathbb{R}^{C \times H \times W} \nonumber \\
    \text{Output:} \quad & \mathbf{Y} \in \mathbb{R}^{H \times W} \nonumber
\end{align}

where $C=5$ (number of seismic shot gather channels), $H=300$ (number of time steps), and $W=1259$ (number of receiver positions), as defined in the ThinkOnward 2025\cite{thinkonward2025} dataset structure.

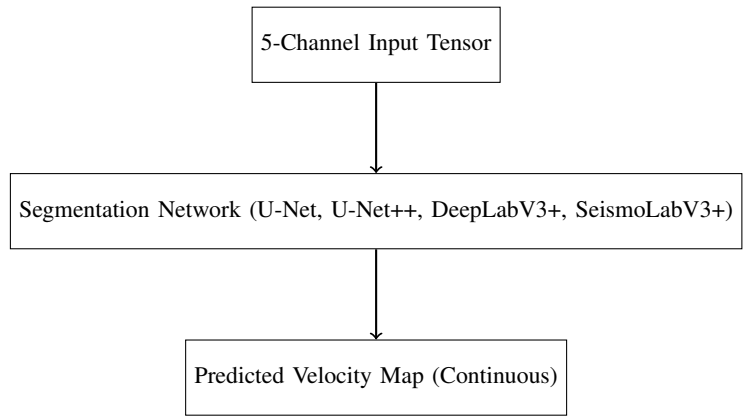
\begin{figure}[htbp]
\centering
\begin{tikzpicture}[node distance=1.2cm, every node/.style={font=\small}]
\node (input) [draw, minimum width=2.8cm, minimum height=1cm] {5-Channel Input Tensor};
\node (model) [draw, below=of input, minimum width=3.5cm, minimum height=1cm] {Segmentation Network (U-Net, U-Net++, DeepLabV3+, SeismoLabV3+)};
\node (output) [draw, below=of model, minimum width=4cm, minimum height=1cm] {Predicted Velocity Map (Continuous)};
\draw[->, thick] (input) -- (model);
\draw[->, thick] (model) -- (output);
\end{tikzpicture}
\caption{Overview of the seismic velocity inversion task as a supervised regression problem. Multi-channel seismic shot gathers are mapped to a continuous-valued velocity field using deep segmentation networks.}
\label{fig:overview}
\end{figure}

In this work, we benchmark multiple deep segmentation networks adapted for this regression task. Each model takes as input the five-channel seismic tensor and outputs a continuous-valued velocity map, as illustrated in Figure \ref{fig:overview}.

\section{Related Research}

Full-waveform inversion (FWI) has long been the benchmark for seismic velocity model building due to its ability to achieve high-resolution reconstructions at the scale of half the propagated wavelength \cite{virieux2009}. 
Despite recent advances in high-performance computing and 3D acoustic applications, its widespread adoption remains limited. FWI suffers from dependence on accurate initial models, sensitivity to noise and frequency limitations, and high computational demands.
Extensions such as reflection-waveform inversion (RWI) and multiparameter inversion have been explored \cite{wang2018demigration}, but remain underdetermined and sensitive to acquisition geometry. These challenges have motivated research into data-driven alternatives.

Deep learning has emerged as a promising approach, reformulating seismic velocity inversion as a supervised regression problem. Convolutional neural networks (CNNs) have demonstrated the ability to directly estimate velocity models from raw seismograms, offering automation and efficiency. However, CNN-based methods are often limited by small receptive fields that capture only local features, overlooking long-range spatial dependencies \cite{wang2023seismic}, \cite{sadad2023efficient}. This has led to an investigation of encoder-decoder architectures, attention mechanisms, and hybrid CNN-transformer frameworks, which aim to better model the global context for improved inversion accuracy. W.~Cao \emph{et al.} (2022) proposed a CNN--LSTM fusion network \cite{wei2021seismic} that jointly extracts spatial and temporal features from seismic gathers. Their method achieved higher inversion accuracy and generalization compared to single-network models, while also providing effective starting models for FWI. But, its performance remains sensitive to training strategies and hyperparameter tuning, with applications largely restricted to RMS and interval velocity estimation. 

F.~Li \emph{et al.} (2022) proposed \cite{li2022deep} a hybrid network (AG-ResUnet)\cite{AG-ResUNet++} that has fully convolutional layers, an attention mechanism, and a residual unit to estimate velocity models from gatherings of the common source point (CSP). Their proposed model improved noise and achieved efficient generalization through transfer learning. However, its performance remains constrained by the scarcity of labeled field data and the absence of physics-guided regularization in complex geological settings.

L.~Mosser \emph{et al.} (2018) utilised the power of generative adversarial networks (GANs) for rapid seismic inversion using domain transfer  \cite{mosser2018rapid}. Enhancing with cycle-consistency, their approach produced velocity fields directly from seismic amplitudes which significantly reduces computational costs compared to FWI. The method demonstrated good performance in terms of noise.  Though It shows promising results on field datasets, such as F3(a 3D seismic survey from the North Sea (Netherlands F3 Block), widely used as a benchmark real-data dataset). But, the accuracy remains limited by the synthetic–to–field domain gaps and the lack of explicit physics-based constraints.

F.~Yang and J.~Ma (2019) proposed a direct mapping using a FCN\cite{long2015fully} (a fully convolutional network designed specifically for sementic segmentation) for the reconstruction of the velocity model directly from raw seismograms \cite{yang2019deep}. Their method demonstrated near-real-time inversion, avoided cycle skipping, and reduced human intervention compared to FWI. 

Despite these advances, most studies evaluate individual architectures in isolation, often under differing datasets and experimental conditions. As a result, a systematic benchmarking of modern segmentation-inspired encoder–decoder networks for seismic velocity inversion remains limited. To address this gap, we adapt and evaluate three advanced architectures: U-Net~\cite{UNet}, U-Net++\cite{unet++}, and SeismoLabV3+ (an optimized variant of DeepLabV3+\cite{Deeplabv3++})—under standardized conditions using the ThinkOnward 2025~\cite{thinkonward2025} dataset. By comparing their regression performance on identical multi-channel inputs and evaluation metrics, this work provides a rigorous assessment of their relative strengths and weaknesses, offering practical insights into the suitability of deep segmentation networks for seismic inversion.

\section{Methodology}
The objective of this work is to map multisource 2D seismic shot gather inputs to their corresponding ground-truth velocity models using supervised learning. This study formulates the task as dense regression with segmentation-style deep neural networks, introducing SeismoLabV3+, an optimized DeepLabV3+ tailored for seismic inversion. To validate this, we benchmark SeismoLabV3+ against U-Net and U-Net++ baselines under standardized preprocessing, hyperparameter tuning, training, and evaluation settings. For all the implementations, we used the Segmentation Models Pytorch library\cite{Iakubovskii:2019}.

\subsection{Dataset Description}
This study uses the official data set provided by the ThinkOnward 2025 Speed \& Structure Challenge\cite{thinkonward2025}. The dataset comprises distinct training and test sets, which facilitates supervised learning and rigorous evaluation. 

\subsubsection{Training Dataset}
The training set consists of 2,000 samples. For every sample, six two-dimensional NumPy array files are provided. 

\begin{itemize}
    \item \textbf{Five input feature files:} These files represent synthetic seismic survey data. These data have been acquired from five different seismic source positions. Each file is named\texttt{receiver\_data\_src\_$<$i$>$.npy}, where $<$i$>$ is one of \{1, 75, 150, 225, 300\}. Each file contains a 2D array of shape $(300, 1259)$, corresponding to 300 time steps and 1,259 receiver positions.
    \item \textbf{One target file:} The \texttt{vp\_model.npy} file contains the ground-truth subsurface velocity model for that sample, also as a 2D array of shape $(300, 1259)$.
\end{itemize}
These paired input and target arrays provide the basis for supervised learning in the inversion of the seismic velocity model.

\subsubsection{Test Dataset}
The test set consists of 150 samples. These are organized in the same manner as the training set. Each test sample resides in a uniquely named folder and includes the five input feature files  where (\texttt{receiver\_data\_src\_$<$i$>$.npy}, for $<$i$>$ in \{1, 75, 150, 225, 300\}), following the same naming convention and data structure as in training. The test set does not include the target file (\texttt{\detokenize{vp_model.npy}}), instead, predictions are submitted to the Thinkonword official site\cite{thinkonward2025} for scoring against hidden ground truth. The accuracy of these predictions is assessed via a public leaderboard ranking showing MAPE score.

\subsection{Data Preprocessing}
 To prepare the data for training and evaluation, the following pre-processing steps were applied consistently across all architectures:

\begin{itemize}
    \item \textbf{Channel stacking} --- the five-shot gathers from source positions are  [1, 75, 150, 225, 300]. These shots were stacked into a 5-channel tensor.
    \item \textbf{Normalization} --- to confirm stabilization of learning, each channel was scaled to the range $[0,1]$.
    \item \textbf{Tensor conversion} --- all data were converted into PyTorch tensors. The shape for inputs is $(C,H,W)$ for inputs and for target is $(H,W)$ 
    \item \textbf{Resizing} --- inputs were interpolated to a uniform resolution of $300 \times 1259$.
    \item \textbf{Train/validation/test splits} ---  For leaderboard scoring, the official challenge splits were preserved so that we can ensure fair comparison across methods. Moreover, for internal evaluation, the 2,000 training samples were divided into 60\% training, 20\% validation, and 20\% test subsets.
\end{itemize}

The technical implementation of the work used segmentation PyTorch model library\cite{Iakubovskii:2019}. Padding have been applied to meet Architecture-specific compatibility for methods.

\begin{itemize}
\item \textbf{U-Net++}: All input files were padded to $320 \times 1280$. The divisibility by 32 ensures compatibility through their decoders.

\item \textbf{U-Net and SeismoLabV3+ (DeepLabV3+ variant)}: All inputs were padded to $304 \times 1264$ which ensures the divisibility by 16 required by their decoders.
 \end{itemize}
The ground-truth velocity models were resized and padded using the same process as the input to ensure spatial alignment. 
\subsection{Network Architectures}
This research benchmarks three deep segmentation networks — U-Net\cite{UNet}, U-Net++\cite{unet++}, and the optimized variant of DeepLabV3+\cite{Deeplabv3++} (SeismoLabV3+) — to assess their effectiveness in inversion of seismic velocity. These models were chosen for their established effectiveness in dense prediction tasks like semantic segmentation and medical image reconstruction, which exhibit structural similarities to seismic inversion. Each model was modified to accept five-channel seismic shot gathers as input and to output continuous-valued velocity maps that are appropriate for geophysical analysis. In this part of the study, we will discuss each architecture principle and its strengths and suitability for the seismic inversion task

\subsubsection{U-Net:}
\begin{figure}[htbp]
    \centering
    \includegraphics[width=\linewidth]{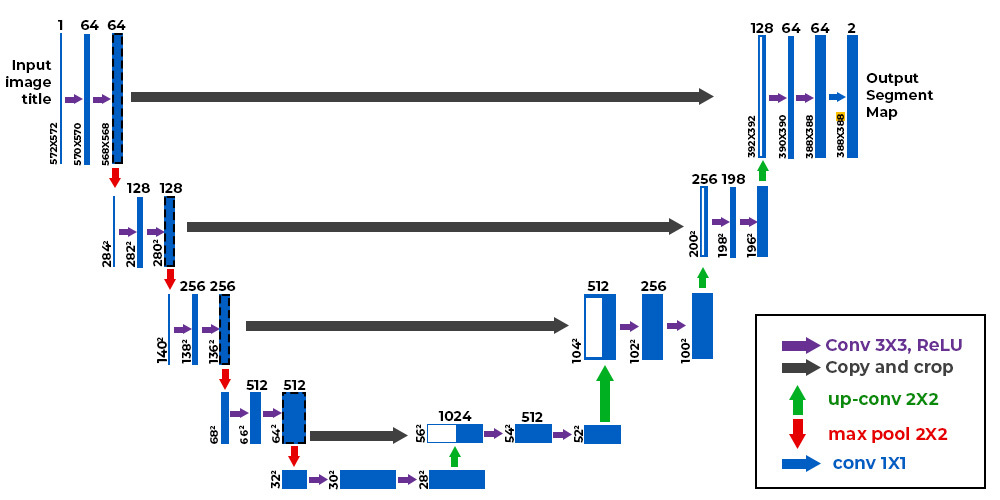}
    \caption{Architecture of U-Net~\cite{UNet}. The model employs a symmetric encoder--decoder structure with skip connections, enabling precise localization by combining high-resolution features from the encoder with decoder upsampling.}
    \label{fig:unet_arch}
\end{figure}

The name “U-Net” originates from the shape of its architecture, which is similar to the English letter “U” as shown in Fig.  ~\ref{fig:unet_arch}. It is widely used in medical imaging because of its superior performance with a limited amount of labeled data. The architecture is symmetric and consists of three key parts, including the Contracting Path (Encoder), which uses small filters (3×3 pixels) to scan the image and identify features. These features are then applied to an activation function called ReLU, adding non-linearity and helping the model learn more effectively. In addition, it uses max pooling (2×2 filters) to shrink the image size while keeping important information. That further helps the network focus on larger features. The bottleneck is in the middle of the “U” where the most compressed and abstract information is stored. It links the encoder and decoder. Expansive Path (Decoder) Uses upsampling, i.e, increasing image size to get back the original image size. This combines information from the encoder using “skip connections.” These connections help the decoder to get spatial details that might have been lost when shrinking the image. It again uses convolution layers to clean up and refine the output\cite{UNet}, \cite{geeksforgeeks_unet}.  

\subsubsection{U-Net++:}
\begin{figure}[htbp]
    \centering
    \includegraphics[width=\linewidth]{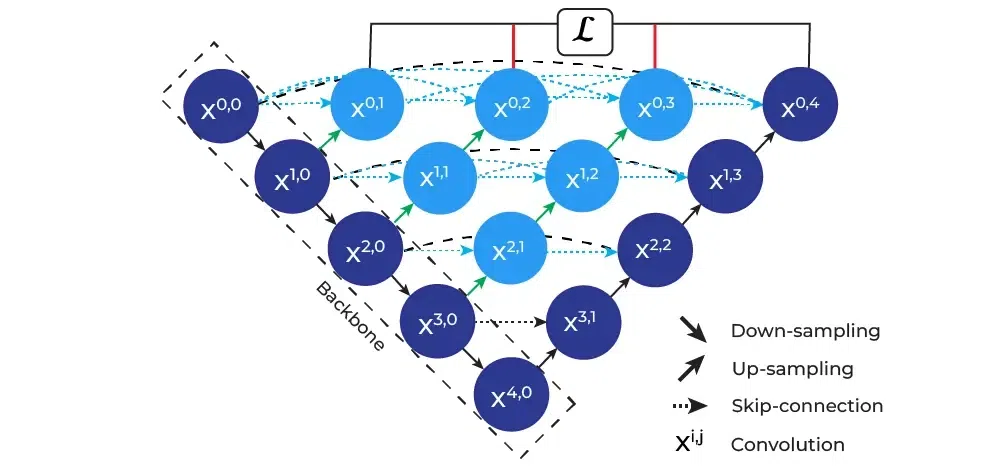}
    \caption{Architecture of U-Net++~\cite{unet++}. This model extends U-Net by introducing nested and dense skip pathways that reduce the semantic gap between encoder and decoder feature maps, thereby improving the recovery of fine structural details.}

    \label{fig:unetpp_arch}
\end{figure}

The U-Net++ architecture is a semantic segmentation architecture based on U-Net, introducing two main innovations in the traditional U-Net architecture. Those are nested dense skip connections and deep supervision, which bridge the semantic gap between encoder and decoder feature maps and improve the gradient flow. Deep supervision improves model performance by providing regularization to the network during training. Fig. ~\ref{fig:unetpp_arch} illustrates the nested encoder and decoder architecture of the U-Net++ architecture. Instead of a traditional skip connection, the lower level feature map is convoluted with the upper-level feature, and then the new combined feature data are passed through \cite{unet++}\cite{geeksforgeeks_unet++}. 

\subsubsection{DeepLabV3 and DeepLabV3+}
Google researchers developed DeepLabV3, a fully Convolutional Neural Network (CNN) model designed to tackle the problem of semantic segmentation. DeepLabV3~\cite{chen2019rethinking} is an incremental update to earlier versions (DeepLabV1~\cite{chen2014semantic} and DeepLabV2~\cite{chen2017deeplab}), and it significantly outperforms its predecessors. 

\begin{figure}[htbp]
    \centering
    \includegraphics[width=\linewidth]{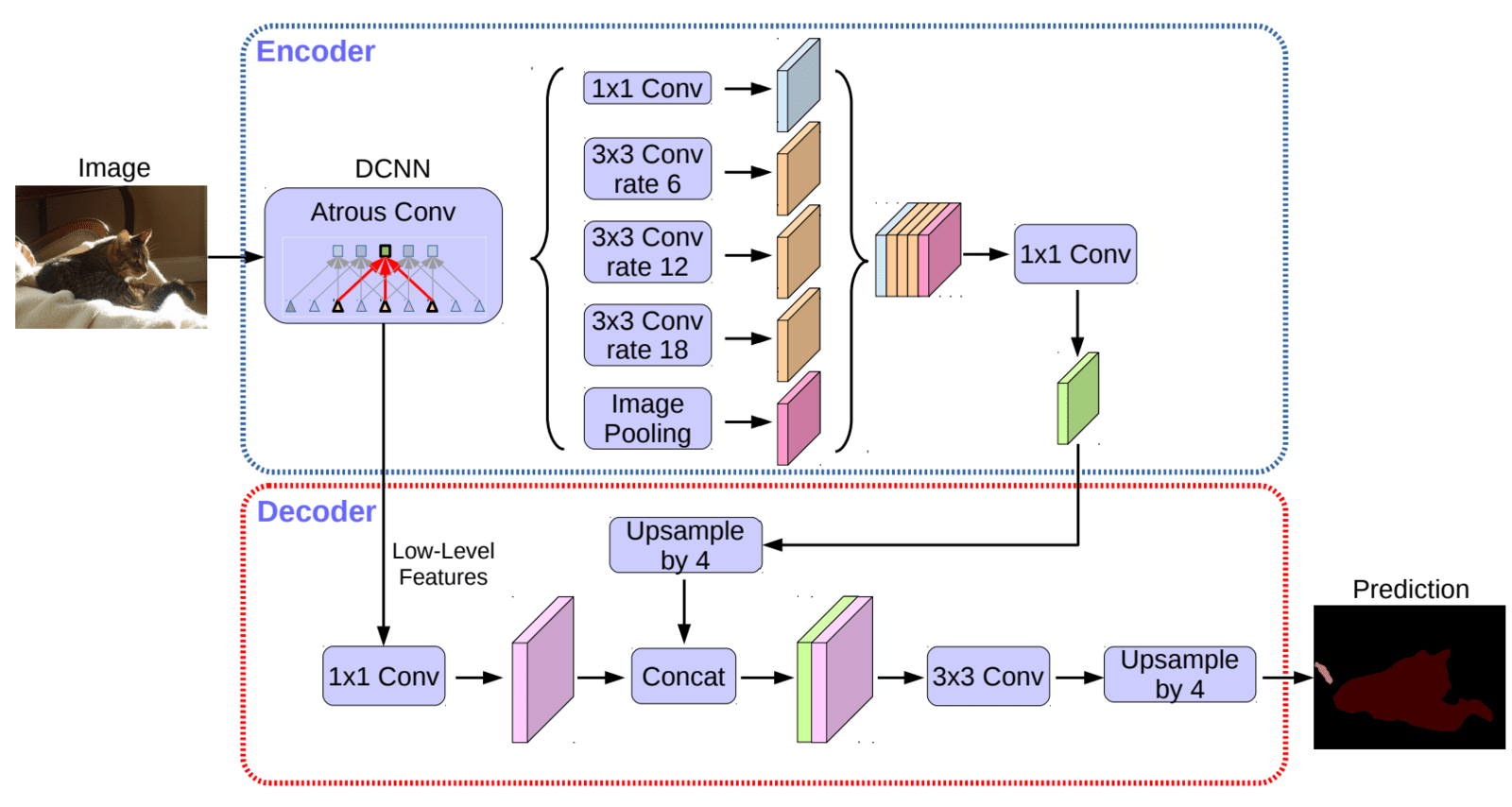}
    \caption{Architecture of DeepLabV3+~\cite{Deeplabv3++}. The model integrates atrous spatial pyramid pooling (ASPP) to capture multi-scale contextual information, and its decoder refines object boundaries by fusing upsampled encoder features with low-level spatial features.}
    \label{fig:deeplabv3p_arch}
\end{figure}

DeepLabV3+~\cite{Deeplabv3++} is an extension of the DeepLabV3 architecture, also developed by Google researchers primarily for semantic segmentation. Figure~\ref{fig:deeplabv3p_arch} illustrates the DeepLabV3+ architecture. DeepLabV3+ utilizes a modified version of the \textit{Aligned Xception} model as its primary backbone (feature extractor) to improve performance with faster computation. In this backbone, each $3 \times 3$ depthwise separable convolution is followed by batch normalization and a ReLU activation. 

The encoder integrates the Atrous Spatial Pyramid Pooling (ASPP) module to capture multi-scale contextual information. The final output of the feature map (before logits) serves as input to the encoder--decoder structure. Within the \textit{encoder}, DeepLabV3+ leverages the ASPP output at an output stride of 16, which provides semantically rich features. Simultaneously, shallow feature maps from earlier layers (containing finer spatial details) are processed using a $1 \times 1$ convolution to reduce dimensionality. 

In the \textit{decoder}, the ASPP features are first bilinearly upsampled by a factor of 4 and then concatenated with the low-level features. This fused representation is refined through successive $3 \times 3$ convolutions to sharpen object boundaries, followed by bilinear upsampling by another factor of 4 to restore the full-resolution segmentation map.

\subsection{SeismoLabV3+ (a task-adapted
variant of DeepLabV3+ optimized for regression with a
resnext50 32x4d backbone)}
SeismoLabV3+ is adaptation of DeepLabV3+~\cite{Deeplabv3++}, tailored specifically for seismic velocity inversion. It incorporates a ResNeXt-50 (32×4d) encoder backbone with ImageNet pretraining, Atrous Spatial Pyramid Pooling (ASPP) with dilation rates (12, 24, 36) for multi-scale context, a decoder refinement module for sharper velocity boundaries, and custom padding to preserve seismic input resolution. DeepLabV3+ itself combines \emph{Atrous Spatial Pyramid Pooling (ASPP)} for multi-scale context with a \emph{decoder refinement module} for boundary sharpening. Our modifications adapt this framework to the unique requirements of seismic inversion.  

The key enhancements of SeismoLabV3 + are summarized as follows:

\begin{itemize}
    \item \textbf{Input adaptation \& regression head:} 
    The first convolutional layer is modified to accept five seismic channels. 
    The classification head is replaced with a single-channel linear layer to predict continuous velocity values. 
    
    \item \textbf{Encoder backbones:} This study evaluated multiple backbone architectures, including ResNet-34, ResNet-50, ResNeXt-50 (32$\times$4d), and EfficientNet-B0, to analyze the trade-offs between accuracy and computational efficiency. Based on these experiments, the default encoder was replaced with ResNeXt-50 (32$\times$4d), which demonstrated comparatively better performance.
    \item \textbf{Pre-trained weights:} 
    Different initialization schemes were tested, including ImageNet, SSL, and SWSL weights. The best results were achieved with ImageNet-pretrained weights.  
    \item \textbf{Hyperparameter tuning:} 
    Each backbone was individually optimized via grid search across learning rate, dropout rate, weight decay, and Atrous Spatial Pyramid Pooling (ASPP) configurations.
\end{itemize}

\subsection{Training Setup}
All experiments  implemented in \texttt{PyTorch}, using the \texttt{segmentation-models-pytorch (SMP)}\cite{Iakubovskii:2019} ensuring consistency across architectures. Training was conducted on NVIDIA A100 GPUs with automatic mixed precision (AMP) enabled to accelerate computation and reduce memory usage in google colab. The same training configuration is used for all models, unless otherwise specified: AdamW optimizer with a weight decay of $1\times10^{-4}$, an initial learning rate of $5\times10^{-4}$, and a \texttt{ReduceLROnPlateau} scheduler to dynamically lower the learning rate on validation plateaus. Each model was trained with 50 epochs along a batch size of 4. A validation split of 20\% of the training data was used with a fixed random seed of 42. The primary loss function was the MAPE (mean absolute percentage error). 

\textbf{U-Net (symmetric encoder–decoder
with skip connections).} The SMP implementation of U-Net was trained using a ResNet-50 encoder (pretrained on ImageNet). The encoder depth was fixed at 5, and the decoder channels were set to $(256,128,64,32,16)$. Input data was padded to $320\times1280$ for divisibility and cropped back to $300\times1259$ in the output layer.

\textbf{U-Net++ (Nested Skip Connections).} U-Net++ was trained with the same ResNet-50 backbone using ImageNet pre-trained weight with the incorporation of nested dense skip pathways to reduce semantic gaps between encoder and decoder features. Decoder channels were identical to U-Net, and inputs were padded to $320\times1280$, with outputs cropped back to $300\times1259$.

\textbf{DeepLabV3+ (Baseline)} To provide a direct baseline for the optimized model, the standard DeepLabV3+ with a ResNet-50 backbone and pre-trained weight ImageNet. This configuration includes the Atrous Spatial Pyramid Pooling (ASPP) module with dilation rates $(12,24,36)$ and a lightweight decoder refinement module that fuses ASPP outputs with low-level encoder features. 

\textbf{SeismoLabV3+ (Optimized Variant).} The proposed SeismoLabV3+ builds upon DeepLabV3+ but introduces task-specific enhancements tailored for seismic inversion: (i) ResNeXt-50 (32$\times$4d) backbone pretrained on ImageNet, (ii) encoder output stride of 16, (iii) ASPP configuration with dilation rates $(12,24,36)$ and dropout probability of 0.5, (iv) single-channel regression head for continuous velocity prediction, and (v) a decoder refinement module that preserves sharp velocity contrasts while maintaining native output resolution ($300\times1259$). This optimized configuration demonstrated superior accuracy in the experiments, achieving the better leaderboard score compared to UNet, UNet++ and baseline deeplabv3+.

\begin{table}[htbp]
\centering
\caption{Training setup and key architectural configurations.}
\resizebox{\columnwidth}{!}{%
\begin{tabular}{|l|l|c|c|c|c|c|}
\hline
\textbf{Model} & \textbf{Backbone} & \textbf{Stride} & \textbf{ASPP} & \textbf{Decoder} & \textbf{Epochs} & \textbf{BS} \\
\hline
U-Net & ResNet-50 & -- & -- & Symmetric & 50 & 4 \\
U-Net++ & ResNet-50 & -- & -- & Nested & 50 & 4 \\
DeepLabV3+ & ResNet-50 & 16 & (12,24,36) & Refinement & 50 & 4 \\
SeismoLabV3+ & ResNeXt-50 & 16 & (12,24,36) & Refinement & 50 & 4 \\
\hline
\end{tabular}%
}
\label{tab:training_setup}
\end{table}


\section{Experiments and Results}
To assess the performance of deep segmentation networks in the context of seismic velocity inversion, we conducted a series of experiments using the  ThinkOnward 2025 Speed \& Structure Challenge~\cite{thinkonward2025} dataset. The primary evaluation metric is used is the Mean Absolute Percentage Error (MAPE). Since the official test set does not include ground-truth velocity models, we designed two experimental setups. In the first, we split the training set into 60\% for training, 20\% for validation, and 20\% for testing Table~\ref{tab:results_internal}. In the second, we followed the official split and reported MAPE scores based on the leaderboard evaluation Table~\ref{tab:results_leaderboard}.

\subsection{Evaluation Metric: MAPE}

The Mean Absolute Percentage Error (MAPE) measures the average magnitude of error between predicted values and actual values, expressed as a percentage. It is defined as:

\begin{equation}
\text{MAPE} = \frac{100\%}{N} \sum_{i=1}^{N} \left| \frac{A_i - F_i}{A_i} \right|,
\end{equation}

where:
\begin{itemize}
  \item \(N\) is the number of data points (fitted points),
  \item \(A_i\) is the actual (ground-truth) value for the \(i\)-th point,
  \item \(F_i\) is the forecast or predicted value for the \(i\)-th point.
\end{itemize}

Lower values of MAPE indicate higher prediction accuracy, with 0\% representing a perfect prediction. This metric is particularly intuitive and useful for comparing models—especially when error interpretation in relative terms is important \cite{MAPE}.

\subsection{Benchmarking Architectures \& Quantitative Analysis}
In the first experiment, the ThinkOnward training dataset ( 2,000 samples ) was divided into 60\% for training, 20\% for validation, and 20\% for testing.  SeismoLabV3+ with a ResNeXt50 backbone achieves the lowest error (0.03025 MAPE). The incremental improvement suggests that each architectural refinement (U-Net → U-Net++ → DeepLabV3+ → SeismoLabV3+) reduces error slightly, and using a stronger backbone (ResNeXt50\_32x4d) gives an additional boost in generalisation. This table ~\ref{tab:results_leaderboard} shows performance on the challenge’s official hidden test set (80\% train, 20\% validation). Errors are slightly higher than those in the internal test split. The difference between internal and leaderboard scores is consistent across models (~0.001–0.003 increase in MAPE).  Again, SeismoLabV3+ is the top performer (0.031246 MAPE) — confirming that the optimised backbone generalises best, even on the official challenge data. 

\begin{table}[htbp]
\centering
\caption{Internal evaluation results (60/20/20 split on training data)}
\resizebox{\columnwidth}{!}{%
\begin{tabular}{|l|l|l|c|}
\hline
\textbf{Model} & \textbf{Backbone} & \textbf{Pre-trained Weights} & \textbf{MAPE (Internal)} $\downarrow$ \\
\hline
U-Net & ResNet-50 & ImageNet & 0.03084 \\
U-Net++ & ResNet-50 & ImageNet & 0.03049 \\
DeepLabV3+ & ResNet-50 & ImageNet & 0.03038 \\
\textbf{SeismoLabV3+} & ResNeXt-50 (32$\times$4d) & ImageNet & \textbf{0.03025} \\
\hline
\end{tabular}%
}
\label{tab:results_internal}
\end{table}

\begin{table}[htbp]
\centering
\caption{Leaderboard evaluation results on the hidden ThinkOnward 2025 test set.}
\resizebox{\columnwidth}{!}{%
\begin{tabular}{|l|l|l|c|}
\hline
\textbf{Model} & \textbf{Backbone} & \textbf{Pre-trained Weights} & \textbf{MAPE (Leaderboard)} $\downarrow$ \\
\hline
U-Net & ResNet-50 & ImageNet & 0.033172 \\
U-Net++ & ResNet-50 & ImageNet & 0.032766 \\
DeepLabV3+ & ResNet-50 & ImageNet & 0.031762 \\
\textbf{SeismoLabV3+} & ResNeXt-50 (32$\times$4d) & ImageNet & \textbf{0.031246} \\
\hline
\end{tabular}%
}
\label{tab:results_leaderboard}
\end{table}


\subsection{Qualitative Analysis of SeismoLabV3+ performance}

\begin{figure}[htbp]
    \centering
    \includegraphics[width=\columnwidth]{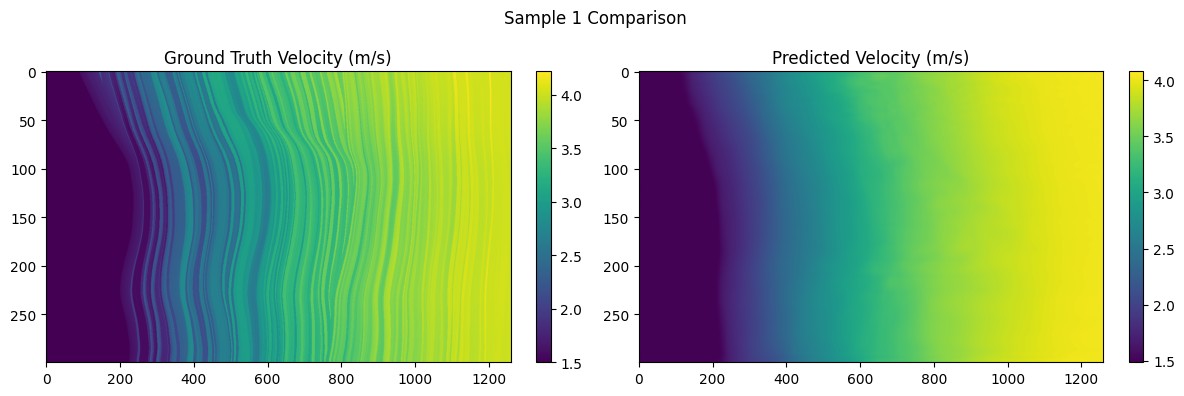}
    \caption{Qualitative comparison of velocity inversion for a random test sample. Left: ground-truth velocity model. Right: predicted velocity model using SeismoLabV3+.}
    \label{fig:qualitative_results}
\end{figure}

Figure~\ref{fig:qualitative_results} presents a qualitative comparison between ground-truth and predicted velocity models in a random test sample from the first experiment. Both figures demontrated the transition from low velocity on the left (blue) to high velocity on the right (yellow). The ground-truth model contains numerous thin, quasi-horizontal bands and laminations, whereas the predicted model appears smoother, with much of this fine stratification absent. While the prediction loses high-frequency details such as sharp layer boundaries, it successfully captures the overall velocity gradient and the low-velocity wedge near x $\ approx$0--300. The alignment of the broad colour trend across both panels indicates that the network effectively models the macro-structure of the subsurface. 

\section{Conclusion}
This research benchmarked multiple deep segmentation networks for seismic velocity inversion using multichannel synthetic seismic shot gathers. This systematic evaluation highlights the potential of encoder–decoder segmentation frameworks for geophysical inversion and emphasises the importance of architectural refinements tailored to the seismic domain. The proposed SeismoLabV3+, a task-adapted variant of DeepLabV3+ modified for regression with a resnext50 32x4d backbone, five-channel seismic inputs, and task-specific optimisations, utilising ImageNet pre-trained weights and carefully tuned hyperparameters, demonstrated superior performance. Future work will extend this study to real seismic field data and explore more sophisticated model architectures for improved inversion accuracy.  


%

\appendices
\section{Reproducibility and Code Availability}
To support reproducibility and further research, the full implementation of all models, preprocessing pipelines, and benchmarking experiments described in this paper has been made publicly available. The source code can be accessed at:

\url{https://github.com/Mahedi-Shuvro/seismic-velocity-inversion}

\section*{Acknowledgment}
The author would like to thank the organisers of the ThinkOnward 2025 Seismic Velocity Inversion Challenge for providing the dataset used in this research.

\ifCLASSOPTIONcaptionsoff
  \newpage
\fi



\bibliographystyle{IEEEtran}
\bibliography{bibtex/bib/IEEEabrv,bibtex/bib/IEEEexample,bibtex/bib/references}

%

\begin{IEEEbiography}[{\includegraphics[width=1in,height=1.25in,clip,keepaspectratio]{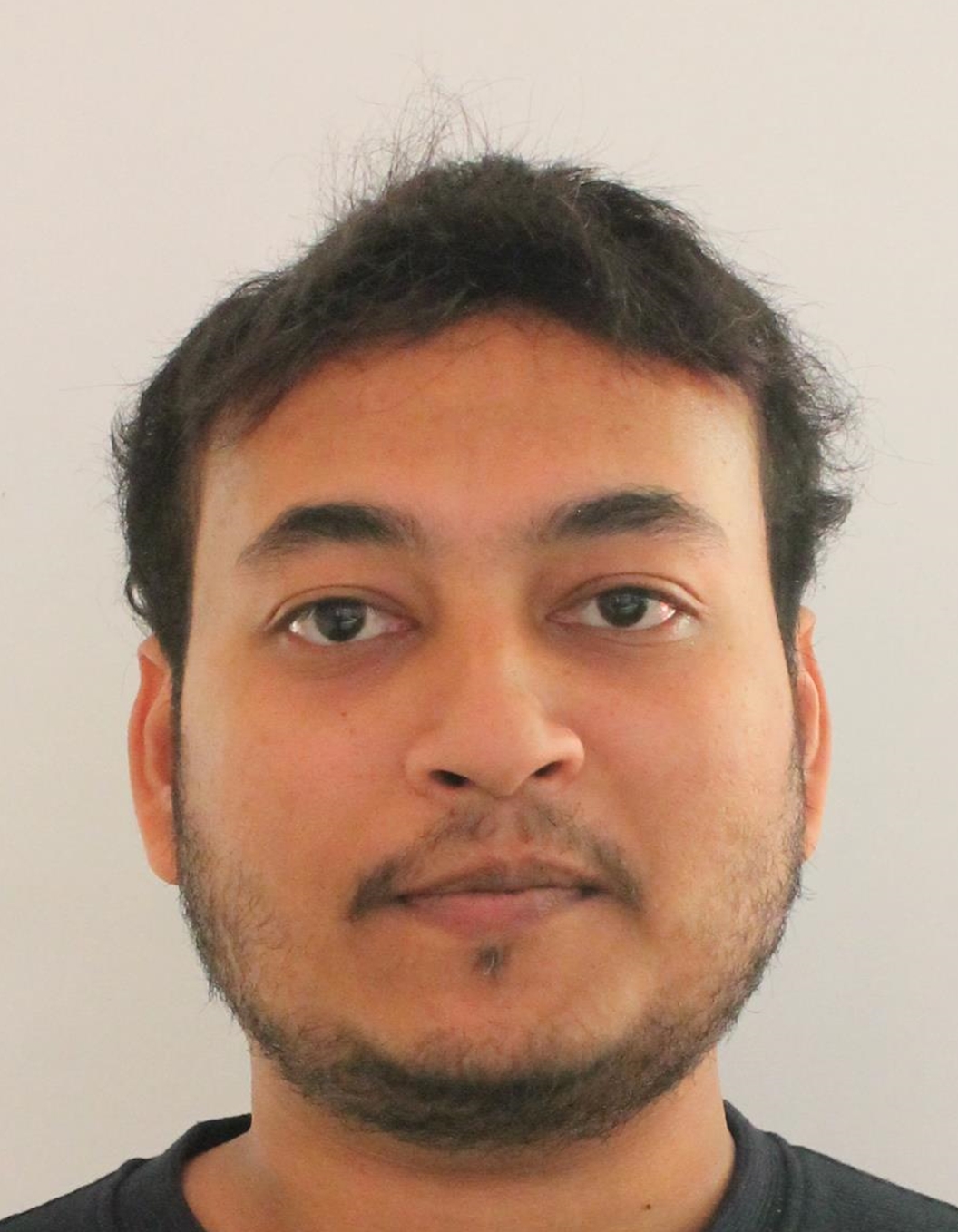}}]{Mahedi Hasan}
received the B.App.Sc. degree in Marine Engineering from the University of Tasmania, Australia, in 2019, and the M.Sc. degree in Artificial Intelligence from RMIT University, Australia, in 2024. His research interests include programming autonomous robots and IoT devices, Machine learning, deep learning, computer vision, AI for maritime and geospatial applications. He is a member of the IEEE and the Australian Computer Society.
\end{IEEEbiography}

\end{document}